\documentclass[11pt,a4paper]{article}

\usepackage{graphicx}
\usepackage{dcolumn}
\usepackage{bm}
\usepackage{ulem}
\usepackage[T1]{fontenc}

\newcommand{\vct}[1]{\boldsymbol{#1}}
\newcommand{\mtrix}[3]{\langle \,#1\,|\,#2\,|\,#3\,\rangle}
\newcommand{\avr}[1]{\langle \,#1\,\rangle}

\begin{document}


\begin{center}
{\Large
The consistent analyses for determination of the point-nucleon distributions
 by electron and proton scatterings
}

\vspace{5mm} 
\noindent
  
Toshio Suzuki$^1$, Rika Danjo$^1$, Toshimi Suda$^1$, 
 Masayuki Matsuzaki$^2$ and Tomotsugu Wakasa$^3$
 
$^1$Research Center for Accelerator and Radioisotope Science,
Tohoku University,\\Sendai 982-0826, Japan
 
 $^2$Department of Physics, Fukuoka University of Education,\\
 Munakata, Fukuoka 811-4192, Japan
 
  $^3$Department of Physics, Kyushu University, Fukuoka 819-0395, Japan

\end{center}



\vspace{3mm}
\noindent
e-mail:kt.suzuki2th@gmail.com

\date{\today}

\begin{abstract}
Electron scattering cross section, as well as proton scattering
cross section, observes the point-proton and the point-neutron
distributions, $\rho_\tau(r), (\tau=p,\, n)$, but both cross sections
are not able to determine them separately.
If they are analyzed consistently with each other,
there is a possibility to determine them with less ambiguity.
The consistency can be examined through the moments of
the charge distribution, $\rho_c(r)$,
which linearly depend on the moments of the point-proton and -neutron
distributions, $\rho_\tau(r),\,(\tau=p,\,n)$. 
The fourth moment, $\avr{r^4}_c$, of $\rho_c(r)$ in $^{208}$Pb
observed in electron scattering is well reproduced by the mean square
radii, $\avr{r^2}_\tau$, of $\rho_\tau(r)$
obtained consistently in the non-relativistic analyses of electron-
and proton-scattering cross sections. 
The regression analyses of the non-relativistic mean-field models
reproduce well those values of the moments.
\end{abstract}



\section{Introduction}\label{intro}

Electron scattering has played an important role for understanding
nuclear structure since the beginning of nuclear physics history \cite{bm}.
The knowledge of the mean square radius (msr)
\footnote{
The abbreviation
of the `rms'(root mean square)-radius is frequently used in the literature,
but it is convenient for the present purpose to employ `msr' for
the mean square radius, because electron scattering observes the value of
the msr, together with the higher moments.}, 
$\avr{r^2}_p$,
of the point-proton distribution, $\rho_p(r)$,
in nuclei is an indispensable information to nuclear physics. 
The reason why $\avr{r^2}_p$ is employed is because it is believed that
the value of $\avr{r^2}_p$ is well determined by the msr, $\avr{r^2}_c$,
of the charge distribution.
The charge distribution, $\rho_c(r)$, is observed with the use of the electromagnetic probes
like electron scattering \cite{vries, ang} and muonic atoms \cite{eut,fri}.
Electromagnetic interaction is well understood theoretically \cite{bd,deforest},
so that the reaction mechanism
is almost completely separated from assumptions on the nuclear structure
which is dominated by strong interaction \cite{deforest}.
As a result, the values of $\avr{r^2}_c$ have been tabulated throughout
the periodic tables \cite{vries,ang,fri}.

The msr, $\avr{r^2}_n$,
 of the point-neutron distribution, $\rho_n(r)$, as the counterpart of $\rho_p(r)$, 
has been studied experimentally through the strong interaction
for a long time, as in Refs.~\cite{ray78,ray79,hoff,star,mark,brown,rh,mh,zen, cry1,cry2}.
It is because $\rho_n(r)$ has no charge and the above electromagnetic probes
interact very weakly with the neutron charge density, $\rho_{cn}(r)$.
In contrast to the electromagnetic interaction, the strong interaction in nuclear medium
are not uniquely understood yet.
Indeed, the above references employ various different parameters and reaction mechanisms
to derive $\rho_n(r)$ from their experiments.
This fact may be a reason why there is no data table which summarized the values of $\avr{r^2}_n$,
as far as the authors know.

Recently, the neutron skin-thickness, $\delta R$, defined by
\begin{equation}
\delta R=\sqrt{\avr{r^2}_n}-\sqrt{\avr{r^2}_p}
\end{equation}
has been widely discussed by using the values of 
$\avr{r^2}_\tau, (\tau=p,\,n)$ derived from the analyses of different probes.
The value of $\delta R$ in $^{208}$Pb is estimated to be about
0.1--0.3 fm \cite{ray78,ray79,hoff,star,mark,brown,rh,mh,zen, cry1,cry2,jlab,di2},
against $\sqrt{\avr{r^2}_c}=5.503$ fm
from electron scattering data \cite{vries, ang}.
When discussing such a small difference of $\delta R$ by using the values derived
in different experiments,
one should analyze experiments consistently by making clear
the definition of $\avr{r^2}_\tau$.
In electron scattering, $\rho_c(r)$ is observed, from which
$\rho_p(r)$ derived in the non-relativistic framework is different from
that in the relativistic one.
 As a result, the values of $\avr{r^n}_p$ obtained from the observed
$\rho_c(r)$ are different in the two frameworks.
In proton scattering, the Lorentz-vector
density, $\rho_{\rm v,\tau}(r)$, used in the analysis with the relativistic
impulse approximation (RIA) is not identical to $\rho_\tau(r)$ used in the
non-relativistic impulse approximation one (NRIA).
The former corresponds to $\rho_\tau(r)$ in the relativistic framework of electron scattering,
while the latter is obtained in the non-relativistic framework of electron scattering.

When the analyses of experiments are performed consistently, then one may compare
their results with those of nuclear models. In that case, the nuclear models should be chosen
which employ the same definitions of 
$\rho_\tau(r)$ and their moments as in the analyses of the experiments.
For example,
among nuclear models,
the mean-field (MF)-models are frequently used, where
there are two model-frameworks. The one is the relativistic (RMF)
and the other the non-relativistic (NRMF)-framework.
In comparing with experiment, the consistent framework should be chosen. 

The MF-models 
 reproduce well gross properties of nuclei as phenomenological models.
They, however, have a set of different interaction parameters from each other
even in the same framework,
according to their own different purpose to explore specific physical quantities.
In comparing to experiment, therefore, it is not appropriate to choose
one model among more than $100$ versions,  accumulated for the last $50$
years \cite{stone}.
Instead of finding one model to reproduce experimental values,
Refs.~\cite{rein0, roca} have proposed to
perform the linear regression analysis (least squares analysis (LSA))
using a set of the MF-models.

At present, the most consistent analyses to determine experimentally
the value of $\delta R$ in $^{208}$Pb
may be performed in electron and proton scatterings,
based on the non-relativistic framework \cite{ray78,ray79,hoff,star,mark},
where the relationship between the moment of $\rho_c(r)$ observed in electron scattering
and those of $\rho_\tau$(r) in proton scattering is clearly defined in the same framework.
The comparison of their results with the NRMF-models are also possible
by using the LSA \cite{kss,ts}.

The purpose of the present paper is to show the consistency between the analyses of electron
and proton scatterings for determination of $\delta R$, and the consistency of the comparison between their results and the NRMF-ones.
In the next section, the definitions of $\rho_c(r)$ , $\rho_\tau(r)$ and their moments
in electron and proton scattering are briefly reviewed.
In Section~\ref{lsa}, the least squares method to analyze the observed moments and
the NRMF-ones in Refs.~\cite{kss,ts} will be mentioned, in particular,
showing the difference
from those in Refs.~\cite{rein0,roca}.
In Section~\ref{dis}, we will discuss the experimental results, comparing to those of NRMF-models
by the LSA.
The final section will be devoted to the brief summary of the present paper.

\section{Electron and proton scattering}\label{el}


Let us briefly review the descriptions of $\rho_c(r)$ and $\avr{r^n}_\tau$
in electron scattering according to Refs.~\cite{kss,ks0,ks1}.
Electron-scattering cross section is analyzed by providing 
the charge distributions, $\rho_{c}(r)$.
The relativistic nuclear charge density is
written as \cite{ks1}
\begin{equation}
 \rho_c(r)
  =\sum_\tau\Bigl( \rho_{c\tau}(r)
  + W_{c\tau}(r) \Bigr),\label{cd2}
\end{equation}
where the proton and neutron charge densities, $\rho_{c\tau}(r)$, are obtained
by convoluting a single-proton and -neutron density, respectively, as 
\begin{eqnarray}
\rho_{c\tau}(r)&=& 
\frac{1}{r}\int_0^\infty dx\, x\rho_\tau(x)\Bigl( g_\tau(|r-x|)
-g_\tau(r+x)\Bigr), \label{cdf}\\[4pt]
W_{c\tau}(r)&=& \frac{1}{r}\int_0^\infty dx\,x W_\tau(x)
\Bigl( f_{2\tau}(|r-x|)-f_{2\tau}(r+x)\Bigr).\label{csof}
\end{eqnarray}
In the above equation, $\rho_\tau(x)$ and $ W_\tau(x)$ stand for the point-nucleon
and point spin-orbit distributions, respectively, and the convolution functions are given by 
\begin{equation}
g_\tau(x)= \frac{1}{2\pi}\int_{-\infty}^\infty dq\, e^{iqx}G_{E\tau}(q^2),
 \hspace{1cm}
f_{2\tau}(x)= \frac{1}{2\pi}\int_{-\infty}^\infty dq\, e^{iqx}F_{2\tau}(q^2),
\end{equation}
where $G_{E\tau}(q^2)$ denote the Sachs form factor, and
$F_{2\tau}(q^2)$ the Pauli form factor \cite{bd}.
In calculating $\rho_c(r)$, we have to choose  
$G_{E\tau}(q^2)$ and $F_{2\tau}(q^2)$ in various estimations 
 in other experiments,  whose detailed discussions
 were given in Refs.~\cite{kss,ks0}.
Ref.~\cite{kss} employed the form factors with the msrs of a single-proton and -neutron
charge distributions to be $r^2_p=(0.877)^2=0.769$ and $r^2_n=-0.116$ fm$^2$, respectively.

The point nucleon density, $\rho_\tau(r)$,
and the spin-orbit one, $W_\tau(r)$, in Eqs. (\ref{cdf}) and (\ref{csof}) are given,
 respectively, by \cite{ks0}
\begin{eqnarray}
\rho_\tau(r)&=& \mtrix{0}{\sum_{k\in\tau} \delta(\vct{r}-\vct{r}_k)}{0},\label{de0}\\
W_\tau(r)&=& \frac{\mu_\tau}{2M}\left(-\frac{1}{2M}\vct{\nabla}^2\rho_\tau(r)
+i\vct{\nabla}\!\cdot\! \mtrix{0}{\sum_{k\in\tau}
\delta(\vct{r}-\vct{r}_k)\vct{\gamma}_k}{0}\right),
\end{eqnarray}
where
the subscript $k$ indicates the nucleon from 1 to $Z$ for $\tau=p$
and to $N$ for $\tau=n$.
Moreover, $M$ denotes the nucleon mass, $939$ MeV, and $\mu_\tau$ the anomalous
magnetic moment to be $\mu_\tau=1.793$ for $p$ and $-1.913$ for $n$. 
The definition of the Dirac matrix, $\vct{\gamma}_k$, is given in 
Ref.~\cite{bd}.
The first equation satisfies $\int d^3r \, \rho_\tau (r) = Z$ for $\tau = p$
and $N$ for $\tau = n$, respectively, while the second equation 
$\int d^3r\, W_\tau(r)=0$, as it should.
Their explicit forms
in the RMF-models are written as \cite{ks0,ks1}
\begin{eqnarray} 
\rho_\tau(r)&=& \sum_{\alpha\in\tau} \frac{2j_\alpha+1}{4\pi r^2}
 \left(G_\alpha^2 + F_\alpha^2\right),\label{de}\\
W_\tau(r)&=& \frac{\mu_\tau}{M}\sum_{\alpha\in\tau} \frac{2j_\alpha+1}{4\pi r^2}\nonumber\\
&\times& \frac{d}{dr}\left(\frac{M-M^*(r)}{M}G_\alpha F_\alpha
+ \frac{\kappa_\alpha +1}{2Mr}G_\alpha^2 -
\frac{\kappa_\alpha - 1}{2Mr}F_\alpha^2\right).
\label{sode}
\end{eqnarray}
In the above equations, $j_\alpha$ denotes the total angular momentum
of a single-particle, $\kappa_\alpha=(-1)^{j_\alpha-\ell_\alpha
 +1/2}(j_\alpha+1/2)$, $\ell_\alpha$ being the orbital angular momentum,
and $M^*(r)$ the nucleon effective mass defined
by $M^*(r)=M+V_\sigma(r)$, where $V_\sigma(r)$ represents the $\sigma$
meson-exchange potential which behaves in the same way
as the nucleon mass in the equation of motion.
The function 
$G_\alpha(r)$ and $F_\alpha(r)$ stand for the radial parts of
the large and small components of the single-particle wave function,
respectively, with the normalization,
\begin{equation}
\int_0^\infty\! dr \left(G_\alpha^2 + F_\alpha^2\right)=1.\label{norm}
\end{equation}
The spin-orbit density
appears, owing to
the anomalous
magnetic moment of the nucleon,
in the relativistic framework,
and its role is enhanced by
the effective mass in relativistic nuclear models.
This enhancement is shown to be necessary for the RMF-models to reproduce
the difference between the charge distributions of $^{48}$Ca and $^{40}$Ca
in Ref.~\cite{ks0}.
The reason why Eq.~(\ref{sode}) is called the spin-orbit density
will be found in Refs.~\cite{ks0,ks1}.

Note that the wave function of the ground-state in Eq.~(\ref{de0})
is defined in the relativistic
framework, as seen in Eqs.~(\ref{de}) and (\ref{sode}). 
Eq.~(\ref{de}) is nothing but the Lorentz vector density, $\rho_{\rm v}(r)$,
used in the RIA-analysis of the proton-scattering cross section \cite{rh,mh}.
The equation in the non-relativistic framework corresponding to Eq.~(\ref{de0})
will be given below.
The spin-orbit density in Eq.~(\ref{sode}) depends not only on 
$\rho_{\rm v}(r)$, but also on the scalar ($G_\alpha^2 + F_\alpha^2$)
and tensor ($G_\alpha F_\alpha$) densities defined
in the RIA of Ref.~\cite{rh}.
Those densities, together with the spin-orbit interaction in the Hamiltonian,
 yield the spin-orbit current through the continuity equation
 of the four-current \cite{nishizaki}.


The mean $2n$th moment $\avr{r^{2n}}_c$ of $\rho_c(r)$ is given by
\begin{equation}
\avr{r^{2n}}_c =\sum_\tau \avr{r^{2n}}_{c\tau},
\qquad
Z\avr{r^{2n}}_{c\tau}=\int d^3r\, r^{2n}\left(\rho_{c\tau}(r)
					  + W_{c\tau}(r) \right).\label{nth1}
\end{equation}
The explicit expressions of  $\avr{r^{2n}}_c$ are provided in Refs.~\cite{kss,ts,ks1}.

Up to this stage, all equations have been given in the relativistic framework.
As far as the authors know, there is no RIA-analysis of the proton-scattering 
cross section
which is as consistent as the NRIA-one
at present \cite{sak2}.
In NRIA, the careful analyses were reported in Refs.~\cite{ray78,ray79,hoff,star,mark}.
They write the optical potential, $U(r)$,  for NRIA as
 \begin{equation}
U(r)=\sum_\tau \int d^3r'\rho_\tau(r)t_\tau(|\vct{r}-\vct{r}'|), \label{ps}
\end{equation}
where $t_\tau$ indicate the nucleon-nucleon $t$-matrix \cite{mark}.
They determined the density distribution, $\rho_\tau(r)$, so as to reproduce
 both electron- and proton-scattering cross sections consistently
 by the iteration method \cite{ray79}, including the relativistic corrections to
 the charge densities \cite{bertozzi}.

The non-relativistic description of $\rho_c(r)$
and the moments with the relativistic corrections
in electron scattering theory are given in Refs.~\cite{kss,ks1}. 
The  description for the two-component wave function in the non-relativistic framework  
is obtained by the Foldy-Wouthuysen (FW) unitary transformation of that for the
four-component one~\cite{bd}.
Because the realistic nuclear Hamiltonian is not known, however,
the previous various authors~\cite{deforest,bertozzi,macvoy}
have used the Dirac equation with electromagnetic field
for the relativistic framework.
In the case of the relativistic Hamiltonian in the $\sigma$-$\omega$ model,
Nishizaki et al. \cite{nishizaki} have obtained the
charge operator $\hat{\rho}(q)$ for
$\tilde{\rho}(q)=\mtrix{0}{\hat{\rho}(q)}{0}$ 
up to order $1/M^{*2}(r)$.
Here, the matrix element
is calculated using the wave functions in the
two component framework, and the operator is written as \cite{ks1} 
\begin{equation}
\hat{\rho}(q) = \sum_{k=1}^Ae^{i\vct{q}\cdot\vct{r_k}}\Bigl( D_{1k}(q^2)
 + iD_{2k}(q^2)\vct{q}\!\cdot\!(\,\vct{p}_k\times\vct{\sigma}_k\,)\Bigr),
\label{fw}
\end{equation}
where $D_1$ and $D_2$ are defined as
\begin{eqnarray}
 D_{1k}(q^2)&=& F_{1k}(q^2)-\frac{q^2}{2}D_{2k}(q^2), \label{d0m}\\[4pt]
 D_{2k}(q^2)&=& \frac{1}{4M^{*2}(r_k)}\left(F_{1k}(q^2)+
 2\mu_k F_{2k}\frac{M^*(r_k)} {M}\right),\label{d2m}
\end{eqnarray}
with the Dirac form factor $F_{1}(q^2)$ related to the Sachs and Pauli form
factors as \cite{bd}
\begin{equation}
F_{1\tau}(q^2)=G_{E\tau}(q^2)+\mu_\tau q^2F_{2\tau}(q^2)/(4M^2).\label{sachs}
\end{equation}
The Fourier transformation of $\tilde{\rho}(q)$
provides the charge distribution in the non-relativistic framework with the
relativistic corrections up to order of $1/M^{*2}(r)$,
\begin{equation}
 \rho_c(r) = \int \frac{d^3q}{(2\pi)^3}\exp(-i\vct{q}\!\cdot\!\vct{r})
 \tilde{\rho}(q).\label{rcd}
\end{equation}
In replacing $M^*(r)$ by $M$, the above equations are the same as
those in Refs.~\cite{deforest, bertozzi, macvoy}.
Thus, the relativistic corrections with $M$ employed in the NRMF-models \cite{sly4} 
are not  equal to those by the RMF-ones with $M^*(r)$. 

The author of Ref.~\cite{ray79} solved Eq.~(\ref{rcd}) with $M^*(r)=M$ to obtain
$\rho_p(r)$, giving the experimental charge density on the left-hand side
and nucleon form factors in the right-hand side.
The point neutron density required in the right-hand side
was given in the iterations from the proton-scattering analyses,
while the spin-orbit density calculated by a one-body potential model
was used \cite{ray79}.

Non-relativistic expressions of the $n$th moment of $\rho_c(r)$ are provided
in Refs.~\cite{kss,ts,ks1,hi}.
In the present paper, we will discuss mainly the 
second (msr) and the fourth moments of $\rho_c(r)$ and $\rho_\tau(r)$.
The non-relativistic expression for the msr
of the above $\rho_c(r)$ is described as
\begin{equation}
\avr{r^2}_{c}
 =\avr{r^2}_p
 +r_p^2+r^2_n\frac{N}{Z}+C_{\rm rel}. \label{correct}
\end{equation}
The relativistic correction, $C_{\rm rel}$, 
up to order of $1/M^{\ast 2}(r)$,
is written as
\begin{equation}
 C_{\rm rel}=\mtrix{0}{\frac{1}{2Z}\sum_{k=1}^A\frac{\mu_k
 \left(2\vct{\ell}_k\!\cdot\!\vct{\sigma}_k + 3(1-M^\ast(r_k)/M)\right)}{MM^\ast(r_k)}
 +\frac{1}{4Z}\sum_{k=1}^Z \frac{2\vct{\ell}_k\!\cdot\!\vct{\sigma}_k +3}
 {M^{\ast 2}(r_k)}}{0}.\label{rr}
\end{equation}
When using the free Dirac equation for the Hamiltonian,
the above relativistic correction is reduced to \cite{kss}
\begin{equation}
C_{\rm rel}= \frac{1}{M^2}\left(\frac{1}{Z}\sum_{k=1}^A\mu_k
 \mtrix{0}{\vct{\ell}_k\!\cdot\!\vct{\sigma}_k}{0}+\frac{3}{4}
 +\frac{1}{2Z}\sum_{k=1}^Z\mtrix{0}
 {\vct{\ell}_k\!\cdot\!\vct{\sigma}_k}{0} \right).\label{r}
\end{equation}
The last term of the right-hand side in the above equation
is obtained in the FW-transformation,
together with the first two terms which have been employed in the literature \cite{sly4}.

The fourth moment of the charge distribution depends not only
on the fourth and the second moments of $\rho_p(r)$ \cite{ks1},
but also on the second moment of $\rho_n(r)$. 
Ref.~\cite{kss} provides $\avr{r^4}_c$ as
\begin{equation}
\avr{r^4}_c =\avr{r^4}_p+\frac{10}{3}r_p^2\avr{r^2}_p
+\frac{10}{3}r_n^2\avr{r^2}_n\frac{N}{Z}+\Delta_4, \label{4thmc}
\end{equation}
where $\Delta_4$ represents the fourth moment of a single proton and neutron
charge distribution and relativistic corrections.
Refs.~\cite{kss,ks1} shows the explicit expression of $\Delta_4$, and its value 
is estimated model-dependently in Ref.~\cite{kss}.
The last three terms of Eq.~(\ref{correct}) for $\avr{r^2}_c$
will be expressed as $\Delta_2$ from now on
in the same way. 
We note that, as discussed in detail in Refs.~\cite{kss,ks1}, 
the relationship between
$\avr{r^n}_c$ and $\avr{r^n}_\tau$ 
in Eqs.~(\ref{correct}) and (\ref{4thmc}) is model-independent,
and should be kept in any estimation of the moments 
in the non-relativistic framework. 
It will be shown in the next section that $\avr{r^2}_n$-dependence of Eq.~(\ref{4thmc}) plays a
role as a bridge between the analyses of electron- and proton-scatterings.

\section{Comparison of the experimental values with those of the nuclear models}\label{lsa}

The experimental values should be
compared with those of the nuclear models in the same framework
as in the analyses of the experiments.
One of the best framework of the phenomenological modes
for heavy nuclei may be the MF-ones.
Among them, the NRMF-models should be used for the present purpose.
We are not interested in individual models in the MF-framework, because
they have different interaction parameters
from each other, which reproduce similarly gross properties
of nuclei \cite{stone}.
Instead, our interest is whether or not the MF-framework has ability
to reproduce the experimental  values  of the various moments.
For this purpose, the analysis using the LSA
employed in Ref.~\cite{kss} is useful.

Let us review the LSA explained in Refs.~\cite{kss,ts}, but in a different  way.
First, the LSA prepares a set, $M$, composed of the MF-models, $m_i$, which are chosen
arbitrarily in the same framework, the  NRMF-framework or the RMF-one.
Second, the reference formula, like Eq.~(\ref{4thmc}), is provided as
\begin{equation}
Y=\sum_{j=1}^N c_jX_j\,,\quad (c_j : {\rm constant}).\label{rf}
 \end{equation}
The value of $Y$ is able to be determined  by experiment, like $\avr{r^4}_c$,
while $X_j$ is its component, like $\avr{r^2}_\tau$, with the constant, $c_j$,
which is definitely given as in Eq.~(\ref{4thmc}).
$N$ denoting the number of the components in the reference formula.
Third, the values of the two correlated variables $X_j$ and $Y$, 
are calculated in each model, $m_i$, as  $(X_{ji},Y_i)$.
Fourth, by plotting the values, $(X_{ji},Y_i)$ in the ($X$--$Y$)-plane,
the linear regression line, which we call the least squares line (LSL), is obtained as
\begin{equation}
Y=a_jX_j+b_j\,, \quad(j=1,2,\cdots N).\label{lsl}
\end{equation}
Fifth, the experimental value of $Y$ is written in the $(X_j, Y)$-plane as
$Y_{\rm exp}=c$,\,($c=$\,constant).
Finally, the cross point of the lines, $Y$and $Y_{\rm exp}$, determines the LSL-value,
$X_{j\rm L}$, for the component, $X_j$, of $Y$. 
 
 The meaning of the LSL-value, $X_{j\rm L}$, is as follows.
On the one hand,
writing the mean value of the results  calculated  
 by the models in the set as
\begin{equation}
<Y_i>=\sum_{j=1}^N c_j<X_{ji}>,\label{rf1}
 \end{equation}
we have
\begin{equation}
Y_i-<Y_i>=\sum_{j=1}^N c_j(X_{ji}-<X_{ji}>).
\end{equation}
On the other hand, the LSL of Eq.~(\ref{lsl}) gives
\begin{equation}
Y_i-<Y_i>= a_j(X_{ji}-<X_{ji}>).
\end{equation}
The above two equation yields a sum rule for the slopes of the LSLs as
\begin{equation}
\sum_{j=1}^N\frac{c_j}{a_j}=1\label{sumrule}
\end{equation}
Now, the LSL-value is defined by
\begin{equation}
Y_{\rm exp}=a_jX_{jL}+b_j,
\end{equation}
which provides
\begin{equation}
<X_{ji}>=X_{jL}-\frac{1}{a_j}(Y_{\rm exp}-<Y_i>).\label{mv}
\end{equation}
Inserting the above equation into Eq.~(\ref{rf1}), using the sum rule, Eq.~(\ref{sumrule}),
we obtain
\begin{equation}
Y_{\rm exp}=\sum_{j=1}^Nc_jX_{jL}.\label{final}
\end{equation}
The expression of Eq.~(\ref{sumrule}) in taking into account the standard deviation of the LSL
is given in Ref.~\cite{ts}.
Thus, the LSA provides uniquely the value of each component of $Y$ by
the LSL-value, $X_{jL}$.
It is clear that the LSL-values are not the experimental values of the components $X_j$,
but the values of the components which the model-framework employed requires
for reproducing the experimental value of $Y$.

For derivation of Eq.~(\ref{final}), the following remarks should be kept in mind.
First, one has to know the reference formula, Eq.~(\ref{rf}), as in Eq.~(\ref{4thmc}),
in the present LSA, in order to choose the variable correlated with experimental value.
Otherwise, even if the LSL is obtained between the two physical quantities,
we can not prove that the LSL-value is the one which is necessary
for reproducing the experimental value as in Eq.~(\ref{final}).
For example, Ref.~\cite{kss} showed the following well defined correlation,
\begin{eqnarray}
\avr{r^2}_c&=&a_{cp}\avr{r^2}_p+b_{cp},\label{cp}\\
 \avr{r^2}_p&=&a_{pn}\avr{r^2}_n+b_{pn},\label{pn}\\
\avr{r^2}_c&=&a_{cn}\avr{r^2}_n+b_{cn}.\label{cn}
\end{eqnarray}
The first equation is a result of the reference formula, Eq.~(\ref{correct}),
while the second equation holds in the MF-framework mainly through
the symmetry- and Coulomb-energy, according to 
the Hugenholtz-Van Hove (HVH) theorem \cite{ts1}.
The third one, which has no reference formula, is due to the first two equations.
If the experimental value of $\avr{r^2}_c$ is given in Eq.~(\ref{cp}),
as one of the input-values for the MF-models, 
then the LSL determines the values  of $\avr{r^2}_p$ and $\avr{r^2}_n$
by the above first two equations.
According to this procedure, it is trivial for the experimental value of $\avr{r^2}_c$
in Eq.~(\ref{cn}) to accept any value of $\avr{r^2}_n$ already determined
 by the first two equations.
Thus, Eq.~(\ref{cn}) does not mean that the experimental value of $\avr{r^2}_c$ 
determines the one of $\avr{r^2}_n$. This fact of Eq.~(\ref{cn})
is called a spurious correlation in Ref.~\cite{ts1}.
The similar discussions were given for the correlation
between $\delta R$ and the slope of the symmetry energy, $L$, in Ref.~\cite{ts1}. 
The reference formula between $\delta R$ and $L$ is not described as in the form of
Eq.~(\ref{rf}) \cite{ts1}.

Second, the set of the models should have the same definition of Eq.~(\ref{rf}).
Hence, for example, NRMF- and RMF-models should not be included in the same set.
Indeed, Refs.~\cite{kss, ts,ts1} show that the NRMF- and RMF-frameworks
yield different LSL-values from each other. The part of those differences
stems from the difference between the reference formulae,
while the other part is due to Eq.~(\ref{pn}) which is different 
between the two-frameworks,
as shown in Ref.~\cite{kss}. 
If the models are mixed in the same set, an unreasonable correlation would appear,
as shown in Ref.~\cite{ts1} in the case of $L$. 

Third, Eq.~(\ref{final}) does not require that the mean value of $Y$
in the set of the models, which are chosen arbitrarily, reproduces its experimental value. 
Moreover, the LSA does not require necessarily
preparing a set by the state of the art models only in the same framework.

One comment should be added to this section.
The above LSA in Refs.~\cite{kss,ts,ts1}
was inspired by Refs.~\cite{rein0,roca}, but 
can not be applicable to the analyses of the correlation between $\delta R$
and the parity violating asymmetry, $A_{\rm PV}$ \cite{jlab,jlab1},
 because there is no reference formula which shows explicitly the relationship
 between $\delta R$ and $A_{\rm PV}$ or $\avr{r^2}_n$ and $A_{\rm PV}$
 in their phase shift analyses of the electron-scattering
 cross section.
Even in the PWBA for the conventional electron scattering,
the form factor squared
is not expressed linearly in terms of $\avr{r^2}_c$.
It is given by \cite{deforest}, 
\begin{eqnarray}
|F_c(q)|^2&=&\sum_{n=0}^\infty(-1)^nq^{2n}\sum_{k=0}^n\frac{\avr{r^{2k}}_c\,
\avr{r^{2(n-k)}}_c}
{(2k+1)!(2n-2k+1)!}\, \nonumber\\
&=&1-\frac{1}{3}q^2\avr{r^2}_c+\frac{1}{180}q^4(3\avr{r^4}_c+5\avr{r^2}^2_c)\nonumber\\
 &&-\frac{1}{2520}q^6(\avr{r^6}_c+7\avr{r^4}_c\avr{r^2}_c) +\cdots \label{fm2}
\end{eqnarray}
which is not a type of Eq.~(\ref{rf}) for $Y=|F_c(q)|^2$ and $X_j=\avr{r^{2j}}_c$.
In order for the second term with $\avr{r^2}_c$ only to dominate the
form factor squared, as in Eq.~(\ref{rf}),
the value of  $q^2$ should be about less than $0.01$ fm$^{-2}$ in $^{208}$Pb,
where the convergence of  the alternating series in Eq.~(\ref{fm2}) is assured
and the reminder term is estimated to be negligible through the Leibniz criteria \cite{hi}. 
The JLab-experiment \cite{jlab} has been performed at $q^2=0.158$ fm$^{-2}$,
where the convergence of Eq.~(\ref{fm2}) as the alternating series is obscure as
\begin{equation}
|F_c(q^2=0.158)|^2=1-1.599+1.129-0.348+\cdots.
\end{equation}
The right-hand side of the above equation
is evaluated, employing the experimental values of $\avr{r^n}_c$ obtained
by the sum-of-Gaussians (SOG)-analyses of the
electron-scattering cross section \cite{kss, ts}. 
If a linear correlation between $|F_c(q)|^2$ and $\avr{r^2}_c$
 is found numerically at a given value of $q^2$ in calculations by the MF-models,
it may be $q$-dependent \cite{ts}, as seen in Eq.~(\ref{fm2}).
In the literature \cite{mf}, it is pointed out that there is 
the disparity between the  $\delta R$-values of $^{208}$Pb
and $^{48}$Ca \cite{jlabca} in the JLab-analyses.
The difference itself between those values, however, is not a problem, because
$\delta R$ has the $I=(N-Z)/A$-dependence
which appears as a result of the HVH theorem in the MF-models \cite{ts1}.
The value of $\delta R$ is larger in $^{208}$Pb than in $^{48}$Ca.
Such a difference has been observed in the LSA in Ref.~\cite{kss}.

In the same way as for $A_{\rm PV}$, there is no reference formula for the relationship
between $\delta R$ and the dipole polarizability, $\alpha_{\rm D}$,
as far as the authors know \cite{di2,rn}.
Note that $A_{\rm PV}$ provides $\delta R= 0.283\pm 0.071$ fm,
while $\alpha_{\rm D}$  $0.156^{+0.025}_{-0.021}$ fm.
If one accepted the LSL-value without the reference formula,
 Eq.~(\ref{cn}) would be enough
 for determining the value of $\avr{r^2}_n$ in the MF-frameworks.
 In fact, such a equation
 was derived in Ref.~\cite{kss},
employing the conventional electron scattering data for $\avr{r^2}_c$ \cite{vries, ang}.  
They obtained $\delta R=0.270$ fm in the RMF-framework,
and $\delta R=0.155$ fm in the NRMF-one, according to the LSA.
Against these values, the LSA with respect to $\avr{r^4}_c$,
according to the reference formula, provides
$\delta R=0.279$ fm in the RMF-framework, and $\delta R=0.160$ fm in the NRMF-one.

\section{Discussions} \label{dis}

\begin{table}
 \hspace*{-0.5cm}%
\begin{tabular}{|l|c|c|c|c|c|c|c|c|} \hline
$\,$&
$\avr{r^2}_p$&
$\avr{r^4}_p$&
$\avr{r^2}_n$&
$\avr{r^4}_n$&
 $\avr{r^2}_c$&
$\avr{r^4}_c$&
$\avr{r^6}_c$&
 $\delta R$
\\ \hline
\rule{0pt}{12pt}%
$(p, p)$  & $29.790$ &$1119.6$&$31.900$&$1317.3$ &$ 30.265$ & $1173.3$ &$\,$ & $0.190$\\
$(e, e)$  & $\, $ &$\,$&$\,$& $ \,$ &$30.283$ & $1171.981$ &$52939.613$&$\,$\\ 
\,LSA(2) & $29.671$ & $\,$ &$\,$& $\,$ &$30.283$&$\,$ &$\,$  & $\,$\\
\,LSA(4) & $29.738$ & $1111.855$ &$31.507$& $\,$ &$\,$&$1171.981$ &$\,$  & $0.160$\\
\,LSA(6) & $29.810$ & $1117.338$ &$31.611$& $1282.926 $ &$\,$&$\,$ &$52939.613$  & $0.163$\\
RLSA(2) & $29.733$ & $\,$ &$\,$& $\,$ &$30.283$&$\,$ &$\,$  & $\,$\\
RLSA(4)   & $29.843$ & $1118.322$ &$32.964$& $\,$ &$\,$&$1171.981$ &$\,$  & $0.279$\\ 
RLSA(6)   & $29.936$ & $1125.605$ &$33.070$& $1408.983\,$ &$\,$&$\,$ &$52939.613\,$ & $0.279$\\ 
$(\bar{p}N)$ & $ (29.554)$ &$(1098.016)$&$31.311$& $\,$ &$\,$ & $(1156.047)$ &$\,$&$0.159$ \\
$(\gamma,\pi^0)$& $(29.569)$ &$(1096.854)$& $31.114$&$\,$ & $\,$& $(1155.040)$ & $\,$&$0.140$\\
 \hline
\end{tabular}
\caption{
The $n$th moments of the charge ($c$), proton ($p$), and neutron ($n$) distribution in $^{208}$Pb obtained
 in various analyses.
The neutron skin-thickness, $\delta R$, is defined by $(\sqrt{\avr{r^2}_n}-\sqrt{\avr{r^2}_p})$.
 The values of $(p,p)$ is taken from the analysis of proton scattering \cite{mark}, while
 those of $(e,e)$ from electron scattering \cite{vries,emrich}.
The LSA(n)-row shows the results of the analysis by the least squares method on the non-relativistic
 mean-field models with respect to the $n$th moment of the charge distribution
 observed in electron scattering \cite{kss,ts}. 
The values of $(\bar{p}N)$ and $(\gamma,\pi^0)$ indicate the experimental values obtained
 in the $\bar{p}N$ \cite{cry1} and $(\gamma,\pi^0)$ \cite{cry2} analyses, respectively.
  The values in the parentheses are calculated with the proton distributions assumed in their
 analyses.  All the values are given in units of fm$^n$. 
   For the details, see the text.  
}
 \label{mom}
\end{table}

The value of $\avr{r^2}_c$ is one of the examples
which are well determined experimentally in nuclear physics,
as used for an input in the MF-models.
Fortunately, $\avr{r^2}_c$ does not depend on the value of $\avr{r^2}_n$.
As a result, the value of $\avr{r^2}_p$ is derived from $\avr{r^2}_c$, but depends
on what kind of the model-framework is employed.
Ref.~\cite{kss} provides $\sqrt{\avr{r^2}_p}$ to be $5.447$ fm
in the NRMF-framework,
while $\sqrt{\avr{r^2}_p}$ to be $5.453$ fm in the RMF-one,
using $\sqrt{\avr{r^2}_c}=5.503$ fm, $r_p=0.877$ fm and $r^2_n=-0.116$ fm$^2$.
Moreover, as mentioned at the end of the previous section,
there is a difference by $0.119$ fm
between the values of $\delta R$ in the NRMF- and the RMF-framework
estimated by LSA.
In determining the small value of $\delta R$,
the analysis of experiment to derive the value of $\avr{r^2}_n$
should be consistent with  that used for $\avr{r^2}_p$.

The author of Ref.~\cite{ray79} aimed to analyze electron and proton
 scattering consistently
for determination of $\rho_n(r)$, employing the following method.
In the first step, the author obtained $\rho_\tau$  
by using experimental values of $\rho_c(r)$ determined
 by electron scattering,
but assuming each contribution
of $\rho_\tau(r)$ to it model-dependently \cite{bertozzi},
because electron scattering can not observe
them separately as mentioned before.
Next, proton scattering is analyzed with the use of the obtained
$\rho_{\tau}(r)$, and the author determined the best $\rho_{\tau}(r)$
to reproduce the proton-scattering cross sections.
Third, the obtained new $\rho_{\tau}(r)$ is examined
if the original electron scattering data are reproduced.
According to such iterations, it is found that a few repetitions are enough
for the convergence, if the first trial function of $\rho_{\tau}(r)$ 
are well prepared \cite{ray79}. The model-dependence in the first step
is expected to disappear in the iterations.

 Such analyses were repeated in Refs.~\cite{ray78,hoff,star,mark} to confirm
 their results.
Nevertheless, even after their studies, investigations of $\avr{r^2}_n$
have still been continued \cite{zen,cry1,cry2,jlab,di2,sak2,jlab1}.
One of the reasons 
why the consistent analyses performed about 30 years ago
were not recognized as a benchmark of the studies on $\avr{r^2}_c$
is because the reaction-mechanism is not uniquely established yet.
The another reason is because 
of the $\rho_n(r)$-profile near the center
which was not well determined 
\cite{hoff,star}, compared with $\rho_p(r)$ derived from
the SOG-analyses of electron scattering \cite{frois}.
This fact implies that 
by comparing $\rho_n(r)$ obtained by one proton-scattering analysis
with others obtained within the proton-scattering ones, we can not
recognize the consistency between analyses of electron- and proton-scatterings.

In noticing
that those proton-scattering analyses do not utilize the shapes  of $\rho_\tau(r)$
as parameters, and keep the consistency of $\rho_\tau(r)$
for reproducing electron- and proton-scattering cross sections, we expect that 
the ambiguity of $\rho_n(r)$ near the center reflects the insensitivity
of proton scattering to the inside of nuclei, but the sensitivity to the nuclear surface
constrained by electron scattering.
According to this speculation, we can use the moments of $\rho_c(r)$
to explore the consistency of the
analyses of the experiments, instead of $\rho_\tau(r)$-profiles.
If $\rho_\tau(r)$ are determined consistently near the surface,
their moments should reproduce $\avr{r^n}_c$ which are a function
of the moments of $\rho_\tau(r)$.
We can expect that
the $\rho_n(r)$-profile near the center
is not important for $\avr{r^2}_n$,  because the moment is given
by $(4\pi/Z)\int dr r^{n+2} \rho_n(r)$.

Fortunately, Ref.~\cite{mark}
summarized their results together with those of Refs.~\cite{ray78,ray79,hoff,star}.
In Table IV of Ref.~\cite{mark},
the values of the $n$th moment
 of $\rho_\tau(r)$ determined by their consistent analyses are listed, where
the values of $\avr{r^n}_c$ observed in electron scattering also listed,
but by assuming the three-point Gaussian distribution for $\rho_c(r)$ in Ref.~\cite{vries2}.
The purpose of Ref.~\cite{mark} published in 1995 was not to reproduce the value of
$\avr{r^n}_c$, according to their analysis of proton scattering,
because the description of $\avr{r^n}_c$ in terms of $\avr{r^n}_\tau$
was not given, until Ref.~\cite{ks1} was published in 2019.

Table \ref{mom} shows the results of Ref.~\cite{mark}, together with other analyses.
The $(p,p)$-row lists their results except for those of the moments of the charge
distribution from electron scattering.
 In order to reproduce 
the values of the $\avr{r^2}_c$ and $\avr{r^4}_c$
using the values of $\avr{r^2}_\tau$  and $\avr{r^4}_p$ in the $(p,p)$-row,
 Eqs.~(\ref{correct}) and (\ref{4thmc}) require $\Delta_2=0.475$ fm$^{2}$
 and $\Delta_4=3.416$ fm$^4$, respectively, as
\begin{eqnarray}
 30.265&=&29.790+0.475,\\
1173.3&=&1119.6+69.400-19.117 +3.416,\label{pps}
\end{eqnarray}
where each value in the right-hand side corresponds to those in Eqs.~(\ref{correct})
and (\ref{4thmc}),
but Ref.~\cite{ray79} cited in Ref.~\cite{mark} used 
the values of $r_p$ and $r_n^2$ to be $0.836$ fm and $-0.117$ fm$^2$, respectively, 
which were taken from Ref.~\cite{holer}. 
These values of $\Delta_n\, (n=2,4)$ are similar to the ones required in the LSA,
as mentioned below.

The experimental values of $\avr{r^n}_c\, (n=2, 4, 6)$ listed in the $(e,e)$-row
are obtained with the use of the charge distribution by
the SOG analysis of electron-scattering cross sections \cite{emrich}.
They are used as the experimental values in the LSA($n$)\, ($n=2, 4, 6$) \cite{kss,ts}
to determine the values of
the corresponding rows in Table~\ref{mom}.
The expression of $\avr{r^6}_c$ in terms
of $\avr{r^n}_\tau$ ($n=2,4,6$ for $p$, $n=2,4$ for $n$)
is given in Refs.~\cite{ts,hi}.
The values of $\Delta_n\, (n=2, 4)$
required to reproduce the experimental values
in the LSA(4)-rows are $0.612$ fm$^2$ and $2.605$ fm$^4$,
respectively, as
\begin{eqnarray}
30.283&=&29.671+0.612,\\
 1171.981&=&1111.855+76.241-18.720+2.605.\label{ees}
\end{eqnarray}

Table~\ref{mom} shows that
the remarkable agreement of the values of the moments in the LSA(4)-row with those
in the $(p,p)$-one,
which are constrained by the value of $\avr{r^4}_c$.
The sum of the first two terms related to $\avr{r^4}_p$ and$\avr{r^2}_p$ as
$\avr{r^4}_p+\frac{10}{3}r_p^2\avr{r^2}_p$ in Eq.~(\ref{4thmc}).
These sums in Eqs.~(\ref{pps}) and (\ref{ees}) become $1189.000$ fm$^4$ and $1188.096$ fm$^4$
in the proton scattering analysis and the LSA(4), respectively.
Thus, it is seen that electron scattering provides a strong constraint on the values of
 the moments of $\rho_p(r)$.

In Table~\ref{mom} are shown the results of the LSA
in the relativistic framework in the RLSA$(n)\,(n=2,4,6)$-rows for reference.
It is seen that the experimental values
of $\avr{r^n}_c\, (n\ge 4)$ play a useful role to explore the consistency
in the experimental determination of $\avr{r^n}_p$ and $\avr{r^n}_n$.  
The $(\bar{p}N)$- and ($\gamma,\pi^0$)-rows list the results of
the analyses of the $\bar{p}\,^{208}$Pb atom \cite{cry1}
and the coherent pion photoproduction \cite{cry2}, respectively.
They assumed the two-point Fermi distributions
describing the point-proton and -neutron densities. The former
obtained the diffuseness parameter, $a_n=0.571$ fm and the half-height radius, $c_n=6.684$ fm,
to reproduce experiment, assuming 
$a_p=0.446$ fm and $c_p=6.684$ fm for the point proton distribution   
which are determined by electron scattering data \cite{fri}.
The latter provides $a_n=0.55$ fm and $c_n=6.70$ fm, using 
$a_p=0.447$ fm and $c_p=6.80$ fm.
The values of the moments, $\avr{r^n}_p$,
calculated using above parameters 
 are given in the parenthesis of the two rows.
 By those values together with $r_p=0.877$ fm and $r^2_n=-0.116$ fm$^2$,
the values of $\avr{r^4}_c$ are obtained as in the parentheses in Table~\ref{mom}.
They are much smaller than the experimental value, implying that consistent analyses
are necessary for discussions of $\avr{r^4}_c$.


We note that the values of the $\avr{r^n}_p$ in the parentheses of Table~\ref{mom} 
are calculated by the following analytic formulae using the Sommerfeld expansion,
instead of the approximate ones used frequently in the literature \cite{bm},
because the exact values of the fourth moments are required for comparison.
For the two-point distribution:
\begin{equation}
\rho_p(r)=\rho_{p0}\left(1+\exp ((r-c_{p})/a_p)\right)^{-1}.
\end{equation}
we have
\begin{equation}
\avr{r^2}_p=\frac{3}{5}c_{p}^2
 \frac{1+\frac{10}{3}(\frac{\pi a_p}{c_{p}})^2+\frac{7}{3}
 (\frac{\pi a_p}{c_{p}})^4}{1+(\frac{\pi a_p}{c_{p}})^2},\label{2pf2}
\end{equation}
\begin{equation}
\avr{r^4}_p=\frac{3}{7}c_{p}^4
 \frac{1+7(\frac{\pi a_p}{c_{p}})^2+\frac{49}{3}
 (\frac{\pi a_p}{c_{p}})^4
 +\frac{31}{3}(\frac{\pi a_p}{c_{p}})^6}
{1+(\frac{\pi a_p}{c_{p}})^2}.\label{2pf4}
\end{equation}



Table~\ref{mom} does not list the errors of the experimental values
and those in the LSAs, because the experimental values are not yet
precise enough to determine the values of the moments quantitatively.
The values of $\delta R$ estimated in each analysis are also listed without errors.
The value of the  $(p,p)$-row is taken from Ref.~\cite{mark}
which did not report the errors, while
Refs.~\cite{ray78,ray79,hoff,star} provide $0.182,0.158,0.14\pm0.04,0.197\pm0.042$
fm, respectively.  
These values may reflect the fact that there remain ambiguities
in their proton-scattering analyses, in addition to the experimental errors.
The experimental value of $\avr{r^2}_c$ in electron scattering has error of
$\pm 0.5\%$ \cite{vries}, while the $\avr{r^4}_c$ $\pm 1.5\%$ \cite{emrich}.
Because of these errors and the standard deviation of the LSL,  
the LSA(4) of the NRMF-models provides $0.162\pm 0.068$ \cite{kss}.
Ref.~\cite{ts} did not estimate the errors in LSA(6), because Ref.~\cite{vries}
did not list enough experimental data for their estimations.
The RLSA(4) yields $\delta R=0.275\pm 0.070$ fm in Ref.~\cite{kss}.
For more precise determination of the experimental values of the moments and
$\delta R$, further investigations are required.



\section{Summary}

In order to obtain the experimental value of $\delta R$
without invoking help of the specific phenomenological nuclear models,
the consistent analyses for determination of the experimental values
of $\avr{r^2}_p$ and $\avr{r^2}_n$ are necessary.
Such analyses of the experiments
are provided for $^{208}$Pb
using electron- and proton-scattering data
in the non-relativistic framework \cite{vries,mark}.
The experimental result is compared with those of
the analyses of the least squares method
on the mean-field models within the same non-relativistic framework \cite{kss,ts}.
The $n$th moments of the charge distribution observed in electron scattering
play a role as a bridge between the analyses of electron- and proton-scattering
for confirming the consistency between them \cite{sds}.
In order to determine the value of $\delta R$, however, it should be explored
if ambiguities 
in proton scattering \cite{ray78,ray79,hoff,star,mark} are reduced more.
In electron scattering also \cite{ds},
the more precise determination of the value of $\avr{r^4}_c$
is necessary for quantitative discussion on $\delta R$.
The consistent analyses of the electron- and proton-scatterings in the relativistic
framework \cite{zen}, together with the relativistic mean-field models,
would improve our understanding  $\delta R$ in nuclei.




%

\section*{Funding}

This work was supported by JSPS KAKENHI Grant Numbers JP22K18706, JP23K25899.

\end{document}